\documentclass[preprint,aps,amssymb,showpacs,superscriptaddress,nofootinbib]{revtex4}
\usepackage{graphicx}
\newcommand{\del}{\partial}
\newcommand{\f}{\frac}
\newcommand{\Case}[2]{{\textstyle \frac{#1}{#2}}}
\newcommand{\lP}{\ell_{\mathrm P}}

\begin{document}
\preprint{IMSc/2007/12/16}

\title{Loop Quantization of Polarized Gowdy Model on $T^3$: Kinematical
States and Constraint Operators}

\author{Kinjal Banerjee}
\email{kinjal@imsc.res.in}
\affiliation{The Institute of Mathematical Sciences\\
CIT Campus, Chennai-600 113, INDIA.}

\author{Ghanashyam Date}
\email{shyam@imsc.res.in}
\affiliation{The Institute of Mathematical Sciences\\
CIT Campus, Chennai-600 113, INDIA.}

\begin{abstract} 
In this second paper on loop quantization of Gowdy model, we introduce
the kinematical Hilbert space on which appropriate holonomies and fluxes
are well represented. The quantization of the volume operator and the
Gauss constraint is straightforward. Imposition of the Gauss constraint
can be done on the kinematical Hilbert space to select subspace of gauge
invariant states. We carry out the quantization of the Hamiltonian
constraint making specific choices. Alternative choices are briefly
discussed. It appears that to get spatial correlations reflected in the
Hamiltonian constraint, one may have to adopt the so called `effective
operator viewpoint'.
\end{abstract}


\maketitle

\section{Introduction}

Classical General Relativity is known to be incomplete and it is widely
believed that its quantum version will address the incompleteness issue.
The main and distinctive feature of general relativity is that the
space-time geometry is dynamical and responsive to matter/energy
distribution. Keeping this feature as central and manifest, a {\em
background independent} quantization strategy has been developed over
the past couple of decades and is referred to as Loop Quantum Gravity
(LQG) \cite{LQGRev,ALReview}. While some of the novel features of this
background independent quantization have been revealed generically, eg
discreteness of {\em Riemannian geometry}, its role in singularity
resolution has been seen only in the simplified homogeneous cosmologies
\cite{LQCRev}. 

These models are obtained by restricting to field configurations
invariant under the action of various groups and go under the name of
{\em symmetry reduced models}.  For the homogeneous models, one is left
with only a finite number of degrees of freedom already at the
kinematical level i.e. before imposition of constraints. The example of
spherically symmetric models have infinitely many degrees of freedom at
the kinematical level but still only a finitely many ones in the
gravitational sector (vacuum model). The next simplest example is that
of polarized Gowdy model on $T^3$ \cite{Gowdy} which has infinitely many
degrees of freedom also at the physical level. As discussed in
\cite{ClassGowdy}, this model has solutions with curvature singularity,
solutions which have bounded curvature but are still ``singular'' in the
sense of having a Cauchy horizon, exhibits a form of BKL scenario, has
an open algebra of constraints and is possibly relevant to see
homogenization at late times. One can expect to learn important lessons
by confronting the background independent quantization strategy to this
model.

A few words on available quantizations may be useful.  These models have
been quantized in the canonical framework in terms of the metric
variables as well as the complex Ashtekar variables.  The first attempts
of quantization, were carried out in ADM variables in
\cite{ADMQuantization}.  Another approach which has been followed is
based on an interesting property of the model. After a suitable
(partial) gauge fixing, these models can be described by (modulo a
remaining global constraint) a ``point particle'' degree of freedom and
by a scalar field $\phi$.  This equivalence was used in the quantization
carried out in \cite{Pierri} and \cite{MenamaruganTorre}. However in
these quantizations, the evolution turned out to be non-unitary and in
\cite{Corichi} a new parametrization was introduced which implemented
unitary evolution in quantum theory.

Canonical quantization of {\em unpolarised} Gowdy $T^3$ model in terms
of the complex Ashtekar variables has been given in \cite{HusainSmolin}
and \cite{Menamarugan}. More recently a hybrid quantization wherein the
homogeneous modes are loop quantized while the inhomogeneous ones are
Fock quantized, has been proposed \cite{MGM}, claiming that loop
quantization of the homogeneous modes suffices to resolve the Gowdy
singularity.

In this paper, we specifically focus on the loop quantization of the
Gowdy model.  The methods and steps used here follow closely those used
in LQG and are to be viewed as first steps towards constructing a
background independent quantum theory of the Gowdy model.  Analogous
steps have been carried out in the case of spherical symmetry
\cite{Spherical1,Spherical2}. Two issues we do not address (but comment
on in the last section) are: (i) viewing Gowdy model as a sector of the
full theory and (ii) retrieving the homogeneous Bianchi I model from
this midisuperspace model. 

The classically reduced Gowdy model has all the ingredients of the full
general relativity: it is a generally covariant field theory on
$\mathbb{R} \times S^1$, its basic fields are 0-forms and connection
1-forms, it has the three sets of first class constraints -- Gauss,
diffeo and Hamiltonian. It is simpler than the full 1 + 3 dimensional
theory in that its graphs will be 1-dimensional, its gauge group is
Abelian ($U(1)$) and flux or triad representation exist (so the volume
operator is simpler). It differs from the full theory in that certain
limits available in the full theory are not available here. For example,
in the full theory one gets back the classical expression of the
constraints in the limit of shrinking the tetrahedra to their base
points (continuum limit). This also shrinks the loops appearing in the
(edge) holonomies, thereby ensuring that the exponents in the holonomies
can be taken to be small. In the reduced theory, however, we have point
holonomies and the exponents are not necessarily small in the continuum
limit. (When the exponents {\em are} components of extrinsic curvature,
they are indeed small in the classical regime as is the case in the
present context.)  Nevertheless, the strategies of background
independent quantization continue to be available and are discussed
below in detail.

In section II, we will introduce the background independent basic
variables and construct the kinematical Hilbert space, define the Volume
operator and solve the Gauss constraint to determine the gauge invariant
spin network states. In section III we will carry out the regularization
of the Hamiltonian constraint. We make specific choices for the
partitions as well as for transcribing the expressions in terms of the
basic variables. Section IV is devoted to the action of the Hamiltonian
constraint on basis states. Section V contains a discussion of
ambiguities in the transcriptions as well as in the choices of
partitions. These have a bearing on incorporating the spatial
correlations in the classical constraint (spatial derivatives) also in
the quantum operator. 

\section{Quantum theory}

\subsection{Preliminary remarks} \label{2a}

In the connection formulation of general relativity, the basic canonical
variables are a real, $SU(2)$ connection $A := A_a^i\tau_i dx^i$ and a
densitized triad $E := \tau^i E^a_i\partial_a$ with the Poisson bracket
given by $\{ A^i_a(x), E^b_j(y) \} = (8 \pi G_{\mathrm{Newton}})\gamma
\delta^b_a \delta^i_j \delta^3(x, y)$. There are three sets of
constraints which can be conveniently presented in matrix notation as
follows. Introduce: 
\begin{eqnarray}
\kappa & := & 8 \pi G_{\mathrm{Newton}} ~ ~,~ ~ \tau_i  ~ ~ :=  -i
\sigma_i/2 \hspace{0.3cm},\hspace{0.3cm} \tau_i\tau_j = -
(1/4)\delta_{ij}\mathbb{I} + (1/2)\epsilon_{ijk}\tau_k, \nonumber
\label{TauDef}\\ 
A_a & := & A_a^i\tau_i \hspace{1.3cm},\hspace{0.3cm} E^a := E^a_i\tau^i
\hspace{0.7cm},\hspace{0.3cm} F_{ab}  := \partial_a A_b - \partial_b A_a
+ [A_a, A_b] \ .
\end{eqnarray}
Then,
\begin{eqnarray}
G(x) & := & G_i\tau^i ~ = ~ \f{1}{\kappa\gamma}\left[\partial_a E^a +
[A_a, E^a] \right] \label{g1} \\
C_a(x) & = & \f{1}{\kappa\gamma}\left[ (-2)\mathrm{Tr}\left(F_{ab} E^b -
A_a G\right)\right] \label{d1} \\
H(x) & = & \f{1}{\kappa}(|\mathrm{det}E^a_i|)^{-1/2}\left[ (-
\mathrm{Tr})(F_{ab}[E^a, E^b]) \right. \nonumber \\ 
& & \left.  \hspace{1.0cm} - 2(1 + \gamma^2) \left(\mathrm{Tr}(E^a
K_a)\mathrm{Tr}(E^bK_b) - \mathrm{Tr}(E^aK_b)
\mathrm{Tr}(E^bK_a)\right)\right] \ . \label{h1}
\end{eqnarray}

For the polarized Gowdy model, the connection and triad variables get
restricted to satisfy $E^x_3 = E^y_3 = E^{\theta}_1 = E^{\theta}_2 = 0$,
$A_{x}^3 = A_y^3 = A_{\theta}^1 = A_{\theta}^2 = 0$. These can then be
expressed in the form \cite{GowdyClassical}:
\begin{eqnarray}
\tau_x(\theta) & := & \mathrm{cos}\beta(\theta) \ \tau_1 +
\mathrm{sin}\beta(\theta) \ \tau_2 \hspace{1.0cm},\hspace{1.0cm}
\tau_y(\theta)  :=  -\mathrm{sin}\beta(\theta) \ \tau_1 +
\mathrm{cos}\beta(\theta) \ \tau_2 \label{tauxtauy}\\
A(\theta) & := & \tau_3 {\cal A}(\theta) d\theta + \left\{\tau_x(\theta)
X(\theta) + \tau_y(\theta) \tilde{X}(\theta)\right\} dx +
\left\{\tau_y(\theta) Y(\theta) + \tau_x(\theta)
\tilde{Y}(\theta)\right\} dy \\
E(\theta) & := & \tau_3 {\cal E}(\theta) \ \partial_{\theta} +
\tau_x(\theta) E^x(\theta) \ \partial_x + \tau_y(\theta) E^y(\theta) \
\partial_y 
\end{eqnarray}

In the above, we have essentially defined $\sum_{i = 1,2}E^x_i\tau^i :=
E^x \tau_x, \sum_{i = 1,2}E^x_i\tau^i := E^y \tau_y$ and demanded that
$\tau_x^2 = -(1/4) = \tau_y^2$. It follows that $[\tau_x, \tau_y] =
\tau_3 $ {\em iff} polarization condition, $\sum_{i = 1,2}E^x_i E^y_i =
0$, holds. This allows us to identify $E^x, E^y$ as the magnitudes of
the two dimensional vectors $\vec{E^x}, \vec{E^y}$ and introduce an
angular coordinate $\beta$ so that $E^x_1 := E^x \mathrm{cos}\beta,
E^x_2 := E^x \mathrm{sin}\beta, E^y_1 := - \mathrm{sin}\beta, E^y_2 :=
E^y \mathrm{cos}\beta$. From these, the definitions of the
$\beta-$dependent $\tau$ matrices follows. The matrices $E^a_i(\theta)$
are now ``diagonal'' for each $\theta$. This fact together with the
properties of $\beta-$dependent $\tau$ matrices, simplifies the
computations. In particular, the co-triad $e$, the spin connection
$\Gamma$ and the extrinsic curvature $K := \gamma^{-1}(A - \Gamma)$ are
obtained as,
\begin{eqnarray}
e & = & \tau_3 \f{\sqrt{E}}{ {\cal E} } d\theta + \f{\sqrt{E}}{E^x}
\tau_x dx + \f{\sqrt{E}}{E^y} \tau_y dy \hspace{1.2cm},\hspace{1.0cm} E
:= E^x E^y |{\cal E}| \\
\Gamma & = & \tau_3 \Gamma^3_{\theta} d\theta + \tau_y \Gamma_x dx +
\tau_x \Gamma_y dy \hspace{3.5cm} \mathrm{where,} \nonumber \\ 
\Gamma_{\theta}^3 & = & - \partial_{\theta} \beta
\hspace{0.3cm},\hspace{0.3cm} 
\Gamma_x ~ := ~ \f{1}{2}\f{E^{\theta}_3}{E^x}
\partial_{\theta}\mathrm{ln}\left(E^{\theta}_3 \f{E^y}{E^x}\right)
\hspace{0.3cm},\hspace{0.3cm} 
\Gamma_y ~ := ~ \f{1}{2}\f{E^{\theta}_3}{E^y}
\partial_{\theta}\mathrm{ln}\left(\f{1}{E^{\theta}_3}
\f{E^y}{E^x}\right) ~ ; \\
\gamma K & = & \tau_3\left( {\cal A} + \partial_{\theta}\beta\right)\ 
d\theta + (\tau_x X  + \tau_y(\tilde{X} - \Gamma_x))\  dx + (\tau_y Y  +
\tau_x(\tilde{Y} - \Gamma_y))\  dy
\end{eqnarray}

The preservation of the polarization condition or equivalently diagonal
form of the extrinsic curvature $K_a^i$ requires, $\tilde{X} = \Gamma_x\
, \ \tilde{Y} = \Gamma_y$.

Thus, the basic variables are $X, Y, {\cal A}, \eta := \beta $ and $E^x,
E^y, {\cal E}, P^{\eta}$ with the Poisson brackets of the form $\{X,
E^x\} = (2 G_{\mathrm{Newton}}/\pi) \gamma \delta(\theta - \theta')$. We
have relabelled $\beta$ by $\eta$ for conformity with the notation of
\cite{GowdyClassical} (modulo a factor of 2). 

Putting $\kappa' := \kappa/(4 \pi^2)$, the constraints take the form,
\begin{eqnarray}
G & := & G_3 ~ = ~ \f{1}{\kappa'\gamma}\left[\partial_{\theta}{\cal E} +
P^{\eta}\right] \label{gauss} \\
C_{\theta} & = & \f{1}{\kappa'\gamma}\left[E^x \partial_{\theta}X + E^y
\partial_{\theta}Y - {\cal A} \partial_{\theta}{\cal E} + P^{\eta}
\partial_{\theta}{\eta} \right] \label{diffeo} \\
H  & = & - \f{1}{\kappa'}\f{1}{\sqrt{E}} \left[ \f{1}{\gamma^2} \left(  X E^x Y E^y + {\cal A} {\cal E}( X E^x  + Y E^y)
+  {\cal E} \partial_\theta \eta (X E^x  + Y E^y) \right) \right.\\
& &  - E^x\Gamma_x E^y\Gamma_y\bigg] \nonumber  + \f{1}{2\kappa'}\del_\theta\left\{\f{2 {\cal E} \left(\del_\theta
{\cal E} \right) }{\sqrt{E}}\right\} 
- \f{\kappa'}{4} \f{G^2}{\sqrt{E}} - \f{\gamma}{2}
\partial_{\theta}\left(\f{G}{\sqrt{E}}\right) \label{hamiltonian}
\end{eqnarray}

It is obvious from these definitions that $X, Y, {\cal E}, \eta$ are
scalars while $E^x, E^y, {\cal A}, P^{\eta}$ are scalar densities of
weight 1.  The Gauss constraint shows that ${\cal A}$ transforms as a
$U(1)$ connection while $\eta$ is {\em translated} by the gauge
parameter. All other variables are gauge invariant.

\subsection{Basic States}

The configuration variable ${\cal A}$ is a $U(1)$ connection 1-form, so
we integrate it along an edge (an arc along the $S^1$) and by taking its
exponential we define the (edge) holonomy variable valued in $U(1)$:
$h^{(k)}_e({\cal A}) := \mathrm{exp}(i \Case{k}{2} \int_e {\cal A} ), k
\in \mathbb{Z}$. The integer label $k$ denotes the representation and
the factor of $1/2$ is introduced for later convenience. The Hilbert
space can be constructed via projective families labelled by closed,
oriented graphs in $S^1$. The graphs are just $n$ arcs with $n$
vertices. Associated with each arc is an edge holonomy in the
representation $k$. For a given graph $\gamma$, consider functions
$\psi$ of $n$ group elements $h^{(k_i)}_{e_i}({\cal A})$ and define an
inner product using the Haar measure on $U(1)$. The projective methods
then allow one to obtain the Hilbert space as a completion of the
projective limits of the graph Hilbert spaces.

The configuration variables $X, Y \in \mathbb{R}$ and $\eta \in
\mathbb{R}/\mathbb{Z}$ are scalars and hence no smearing is needed. For
these we define the point holonomies (at points $v$): $h_v^{(\mu)}(X) :=
\mathrm{exp}(i\Case{\mu}{2} X(v)), h_v^{(\nu)}(Y) :=
\mathrm{exp}(i\Case{\nu}{2} Y(v)), h_v^{\lambda}(\eta) :=
\mathrm{exp}(i\lambda \eta(v))$. Again, the factor of $1/2$ is
introduced for later convenience. A similar factor is {\em not}
introduced for the $\eta$ holonomy since $\eta$ is already an angle. The
$X, Y$ point holonomies are interpreted as unitary representation of the
compact, Abelian group $\mathbb{R}_{\mathrm{Bohr}}$ which is the Bohr
compactification of the additive group of real numbers, $\mathbb{R}$
\footnote{The functions \{$\mathrm{exp}(i \mu X), \mu \in \mathbb{R}$\}
form a separating set of functions to separate points in $\mathbb{R}$.
These are also characters of the topological group $\mathbb{R}$.  Their
finite linear combinations give almost periodic functions of $X$. From
these one constructs a commutative C* algebra with unity. The spectrum
of this algebra happens to be the Bohr compactification,
$\mathbb{R}_{\mathrm{Bohr}}$, of the topological group $\mathbb{R}$.
Its (unitary) irreducible representations are one dimensional and are
labelled by real numbers.  The point holonomies are the representatives.
The Haar measure on this compact group can be presented as: $\lim_{T \to
\infty} \Case{1}{2T}\int_{-T}^T$. With this measure, the Hilbert space
of functions on the group is defined via the inner product: $\langle f,
g\rangle := \lim_{T \to \infty} \Case{1}{2T}\int_{-T}^T dX f^*(X)g(X)$. }.
The representation labels $\mu, \nu$ take values in $\mathbb{R}$.  By
contrast, $\eta$ is an angle variable, so the corresponding point
holonomy is valued in $U(1)$. The representation label then takes only
integer values, $\lambda \in \mathbb{Z}$. The corresponding Hilbert
spaces are constructed again via projective families -- now labelled by
finite sets of points which can be taken to be the vertices of the
graphs used in the previous paragraph.

The kinematical Hilbert space for the model is thus a tensor product of
the Hilbert spaces constructed for ${\cal A}, X, Y, \eta$ variables. A
convenient orthonormal basis for this is provided by the ``charge
network functions'' which are labelled by a close, oriented graph $G$
with $n$ edges $e$ and $n$ vertices $v$, a $U(1)$ representation $k_e$
for each edge, a $U(1)$ representation $\lambda_v \in \mathbb{Z}$ for
each vertex and $\mathbb{R}_{\mathrm{Bohr}}$ representations $\mu_v,
\nu_v$ for each vertex:
\begin{eqnarray} \label{BasisStates}
T_{G, \vec k, \vec \mu, \vec \nu, \vec \lambda}({\cal A}, X, Y, \eta)
&:=& \prod_{e\in G} k_e(h^{(e)})~ \prod_{v \in {V}(G)} \mu_v (h_v(X))
\nu_v (h_v(Y)) \lambda_v (h_v(\eta)) \\
&=& \prod_{e\in G} \exp \left(i \Case{k_e}{2} \int_e {\cal A} \right)
\prod_{v \in {V}(G)}\Big(\exp \left(i \Case{\mu_v}{2} X \right) \exp
\left(i \Case{\nu_v}{2} Y \right) \exp \left( i \lambda_v \eta \right)
\Big)\label{spinstate1} \nonumber
\end{eqnarray}
where $V(G)$ represents the set of vertices belonging to the graph $G$.
The functions with any of the labels different, are orthogonal -- in
particular two graphs must coincide for non-zero inner product.

These basis states provide an orthogonal decomposition for the
kinematical Hilbert space when all the representation labels are
non-zero. 
\subsection{Flux Operators} \label{flux}
The conjugate variables are represented as $E^x (\theta) \sim -i \gamma
\lP^2 \f{\delta h_{\theta}(X)}{\delta X(\theta)}\f{\partial}{\partial
h_{\theta}(X)}$, where $\lP^2 := \kappa'\hbar$.  The flux variables
corresponding to $E^x, E^y, P^{\eta}$ are defined by integrating these
densities on an interval ${\cal I}$ of the circle, eg ${\cal F}_{x,{\cal
I}} := \int_{\cal I} E^x, {\cal F}_{y,{\cal I}} := \int_{\cal I} E^y$.
${\cal E}$ being a scalar, is already a suitable variable.  Their actions
on the basis functions (\ref{BasisStates}) are:
\begin{eqnarray}
\hat{\cal E}(\theta) T_{G, k,\mu,\nu,\lambda} & = & \f{\gamma\lP^2}{2} ~
\f{k_{e^+(\theta)} + k_{e^-(\theta)}}{2} T_{G, k,\mu,\nu,\lambda} \\
\int_I \hat{E}^x T_{G, k,\mu,\nu,\lambda} & = & \f{\gamma\lP^2}{2}
\sum_{v \in V(G) \cap {\cal I}} \mu_v T_{G, k,\mu,\nu,\lambda} \\
\int_I \hat{E}^y T_{G, k,\mu,\nu,\lambda} & = & \f{\gamma\lP^2}{2}
\sum_{v \in V(G) \cap {\cal I}} \nu_v T_{G, k,\mu,\nu,\lambda} \\
\int_I \hat{P}^{\eta} T_{G, k,\mu,\nu,\lambda} & = & \gamma\lP^2 \sum_{v
\in V(G) \cap {\cal I}} \lambda_v T_{G, k,\mu,\nu,\lambda} 
\end{eqnarray}
where ${\cal I}$ is an interval on $S^1$. The $e^{\pm}(\theta)$
refer to the two oriented edges of the graph G, meeting at $\theta$ if
there is a vertex at $\theta$ or denote two parts of the same edge if
there is no vertex at $\theta$. The $k$ labels in such a case are the
same. The $\cap$ allows the case where a vertex may be an end-point of
the interval ${\cal I}$. In such a case, there is an additional factor
of $\Case{1}{2}$ for its contribution to the sum. This follows because
\begin{eqnarray} \int_a^b dx \delta(x - x_0) & = & \left\{
\begin{tabular}{lcl} 1 & ~  if ~ & $x_0 ~ \in ~ (a, b)$ ; \\ $\f{1}{2}$
& ~  if ~ & $ ~ x_0 = ~ a ~ \mathrm{or} ~ x_0 ~ = ~ b $; \\ 0 &  ~ if~
& $ x_0 ~ \notin ~ [a, b] $.  \end{tabular} \right.  \end{eqnarray}
Note that classically the triad components, $E^x, E^y$ are positive.
The fluxes however can take both signs since they involve integrals
which depend on the orientation. 

This completes the specification of the kinematical Hilbert space
together with the representation of the basic background independent
variables. Next we turn to the construction of certain operators.

\subsection{Construction of more general Operators}

The diffeomorphism covariance requires all operators of interest
(constraints and observables) are integrals of expressions in terms of
the basic operators. Secondly, operators of interests also involve
products of elementary operators at the same point (same $\theta$) and
thus need a ``regularization''. As in LQG, the general strategy to
define such operators is: (i) replace the integral by a Riemann sum
using a ``cell-decomposition'' (or partition) of $S^1$; (ii) for each
cell, define a regulated expression choosing suitable ordering of the
basic operators and evaluate the action on basis states; (iii) check
``cylindrical consistency'' of this action in (ii) so that the
(regulated) operator can be densely defined on the kinematical Hilbert
space via projective limit; (iv) finally one would like to remove the
regulator.  One would like to do this in such a manner that the
constructed limiting operator has the same properties under the
diffeomorphism. To achieve this,  usually one has to restrict the
cell-decomposition in relation to a graph. 

In the present case of one dimensional spatial manifold, both the
cell-decomposition and the graphs underlying the basis states are
characterised by finitely many points and the arcs connecting the
consecutive points.  As in LQG \cite{ALReview}, the products of
elementary variables are regulated by using a ``point splitting'' and
then expressing the fields in terms of the appropriate holonomies and
fluxes both of which need at most edges and at each point there are
precisely two edges (in the full theory one needs edges as well as close
loops and there can be an arbitrary number of these). A regulator, for
each given graph G, then consists of a family of partitions,
$\Pi^{G}_{\epsilon}$, such that for each $\epsilon$, the partition is
such that each vertex of $G$ is contained in exactly one
cell\footnote{This is one possible natural choice of a class of
partitions adapted to a graph. The vertices of $G$ always lie in the
interior of the cells and some cells have no vertices. We discuss this
further in the last section.}. There is also a choice of representation
labels $k_0, \mu_0, \nu_0, \lambda_0 $ made which can be taken to be the
same for all $\epsilon$. The regulated expressions constructed depend on
$\epsilon$ and are such that one recovers the classical expressions in
the limit of removing the regulator ($\epsilon \to 0$). There are of
course infinitely many such regulators. A diffeomorphism covariant
regulator is one such that if under a diffeomorphism the graph $G \to
G'$, then the corresponding partitions also transform similarly. Since
each $\Pi_G$ can also be thought of as being defined by a set of points
such that each vertex is flanked by two points (between two consecutive
points, there need not be any vertex), any orientation preserving
diffeomorphism will automatically preserve the order of the vertices and
cell boundaries. Every sufficiently refined partition then automatically
becomes a diffeomorphism covariant regulator. This is assumed in the
following.

As in LQG, the issue of cylindrical consistency is automatically sorted
out by referring to the orthogonal decomposition of ${\cal
H}_{\mathrm{kin}}$ i.e. by specifying the action of the operators on
basis states with all representation labels being non-zero. 

With these preliminaries, we proceed to define some operators.

\subsection{Volume Operator}
In the classical expression for the Hamiltonian constraint, powers of $E
:= |{\cal E}|E^x E^y$, occur in the same manner as in the full theory.
It is therefore natural to consider the expression for the volume of a
region ${\cal I}\times T^2$ and construct the corresponding operator.
With the canonical variables chosen, the volume involves only the
conjugate momenta whose quantization is already done.  The volume
operator written in terms of the basic operators :
\begin{eqnarray}
\mathcal{V} (\mathcal{I} \times T^2)& = &\int_{\mathcal{I} \times T^2}
d^3 x \sqrt {g} \nonumber \\
&=& 4 \pi ^2 \int_\mathcal{I} d\theta \sqrt { | {\cal E} |  E^x
E^y}\label{volumeexpression} \nonumber \\ 
\mathcal{V}_{\epsilon} (\mathcal{I}) & \approx &  \sum_{i = 1}^n
\int_{\theta_i}^{\theta_{i} + \epsilon} \sqrt{|\mathcal{E}|E^x E^y}
\nonumber \\
& \approx &  \sum_{i = 1}^n \epsilon \sqrt{|\mathcal{E}(\bar \theta_i)|}
\sqrt{E^x(\bar \theta_i)} \sqrt{E^y(\bar \theta_i)} \nonumber \\
& \approx &  \sum_{i = 1}^n \sqrt{|\mathcal{E}|} \sqrt{\epsilon E^x}
\sqrt{\epsilon E^y} \nonumber \\
& \approx &  \sum_{i = 1}^n \sqrt{|\mathcal{E}|(\bar \theta_i)}
\sqrt{\left|\int_{\theta_i}^{\theta_i + \epsilon} E^x\right|}
\sqrt{\left|\int_{\theta_i}^{\theta_i + \epsilon} E^y\right|} 
\end{eqnarray}
The right hand side is expressed in terms of flux variables, so the
regulated volume operator can be defined as: 
\begin{equation}
\hat \mathcal{V}_{\epsilon} (\mathcal{I}) :=  \sum_{i = 1}^n \sqrt{\hat
{|\mathcal{E}|}(\bar \theta_i)} \sqrt{\widehat {\left|\int_{ {\cal I}_i}
E^x\right|} } \sqrt{\widehat {\left|\int_{ {\cal I}_i} E^y\right| } }
\label{volumeoperator}
\end{equation}
Clearly this is diagonal in the basis states. and its action on a basis
state $T_{G, \vec{k}, \vec{\mu}, \vec{\nu}, \vec{\lambda} }$ gives the
eigenvalue,
\begin{equation}
V_{ \vec k, \vec \mu, \vec \nu, \vec \lambda} =  \f{1}{\sqrt{2}}
\left(\f{\gamma l_P^2}{2}\right)^{3/2}\sum_{v \in {\cal I} \cap V(G)}
\bigg(|\mu_v|\ |\nu_v|\ | k_{e^{+}(v)} +
k_{e^{-}(v)}|\bigg)^{\f{1}{2}}\label{volumeeigenvalues}
\end{equation}

{\em Remarks}: 

(1) In the above, ${\cal I}_i$ denotes the $i^{\rm {th}}$ cell of the
partition and $\bar \theta_i$ denotes a point in that cell -- it need
not be an end-point of the interval. We have also assumed the ``length
of the intervals'' to be same and equal to $\epsilon$. This corresponds
to a ``cubic'' partition and is chosen for convenience only. We will
always use such partitions in all the operators below.

(2) Although we could restrict to $\mu_v, \nu_v > 0$, it will be more
convenient (eg in the Hamiltonian constraint below) to allow both signs
(corresponding to the orientation of the interval). The eigenvalues of
the volume operator then must have explicit absolute values. We have
thus used the absolute value operators defined from the flux operators. 

(3) For a given graph, the partition (of $\mathcal{I}$) is so chosen
that each vertex is included in one and only one interval ${\cal I}_i$.
For those intervals which contain no vertex of the graph, there is no
contribution to the summation since flux operators have this property.
Hence, the sum collapses to contributions only from the vertices,
independent of the partition. The action is manifestly independent of
$\epsilon$ and even though the number of intervals go to infinity as
$\epsilon \to 0$, the action remains {\em finite} and well defined. 

Because of this property of the fluxes, we can choose the $\bar
\theta_i$ point in a cell to coincide with a vertex of a graph if ${\cal
I}_i$ contains a vertex or an arbitrary point if ${\cal I}_i$ does not
contain a vertex. Such a choice will be understood in the following.

(4) For intervals $\mathcal{I} \neq S^1$, a choice of diffeo-covariant
regulator retains the $v \in \mathcal{I}\cap V(G)$ and hence the action
is diffeo-invariant. The eigenvalues are also manifestly independent of
``location labels'' of the states. For the total volume (which is
diffeo-invariant), the operator is manifestly diffeo-invariant.

\subsection{Gauss Constraint}

Consider the Gauss constraint (\ref{gauss}):
\begin{eqnarray}
G_3 &=& \int_{S^1} d\theta (\del_{\theta} {\cal E} +  P^{\eta})
\nonumber\\ 
& \approx  & \sum_{i = 1}^n \int_{\theta_i}^{\theta_i + \epsilon}
(\del_{\theta} {\cal E} +  P^{\eta}) \nonumber \\
& \approx  & \sum_{i = 1}^n \left[\int_{ {\cal I}_i} P^{\eta}  + {\cal
E}(\theta_i + \epsilon) - \mathcal{E}(\theta_i)\right] \\
\hat G_3^{\epsilon} & :=  & \sum_{i = 1}^n \left[\widehat{\int_{ {\cal
I}_i } P^{\eta}}  + \hat {\cal E}(\theta_i + \epsilon) - \hat
\mathcal{E}(\theta_i)\right] 
\end{eqnarray}

Again, this is easily quantized with its action on a basis state $T_{G,
\vec{k}, \vec{\mu}, \vec{\nu}, \vec \lambda}$ giving the eigenvalue, 
\begin{eqnarray}
{\gamma l_P^2}\sum_{v \in V(G)} \left[ \lambda_v + \f{k_{e^{+}}(v) -
k_{e^{-}}(v)}{2}\right] \label{quantumgauss}
\end{eqnarray}
Notice that in the limit of infinitely fine partitions, for a given
graph, if there is a vertex $v \in {\cal I}_i$, then there is {\em no
vertex} in the adjacent cells. As a result, ${\cal E}(\theta_{i + 1})$
gives the $k_{e^+}(v)/2$ and $- {\cal E}(\theta_i)$ gives $-
k_{e^-}(v)/2$, since $\theta_i$ divides the same edge and so does
$\theta_{i + 1}$.

Once again, the eigenvalues are manifestly independent of $\epsilon$ and
the action is diffeo-invariant. Imposition of Gauss constraint can be
done simply by restriction to basis states with labels satisfying
$\lambda_v = - (k_{e^+(v)} - k_{e^-(v)})/2 , \forall v \in V(G)$. Since
$\lambda_v \in \mathbb{Z}$, the difference in the $k$ labels at each
vertex must be an {\em even} integer. We will assume these restrictions
on the representation labels and from now on deal with gauge invariant
basis states. The label $\vec{\lambda}$ will be suppressed and terms
proportional to the Gauss law constraint in the Hamiltonian constraint
will also be dropped.

Substituting for $\lambda_v$ for each of the vertices and rearranging
the holonomy factors, one can write the gauge invariant basis states are
explicitly given by, 
\begin{eqnarray}
T_{G,\vec k,\vec \mu,\vec \nu}= \prod_{e\in G} \exp \left\{ i
\Case{k_e}{2} \int_e \left( {\cal A} (\theta) - \del_{\theta} \eta
\right) \right\} ~ \prod_{v \in {V}(G)}\Big( \exp \left\{i
\Case{\mu_v}{2} X(v) \right\} ~ \exp \left\{ i \Case{\nu_v}{2} Y(v)
\right\}  \Big) \ . \label{spinstate2}
\end{eqnarray}
We have also used, $\eta(v^+(e)) - \eta(v^-(e)) = \int_e \del_{\theta}
\eta$,  where $v^{\pm}(e)$ denote the tip and tail of the edge $e$.

\section{Hamiltonian Constraint}

The Hamiltonian constraint is a more complicated object. Let us write
(\ref{hamiltonian}) as a sum of a kinetic term and a potential term, 
\begin{eqnarray}
H  & := & - \f{1}{\kappa'}[ H_K + H_P ] \\
H_K & := & \f{1}{\gamma^2} \int_{S^1} \mbox{d} \theta N (\theta)
\f{1}{\sqrt{E}} \left[  X E^x Y E^y + \left({\cal A} +
\partial_{\theta}\eta\right) {\cal E}( X E^x  + Y E^y)  \right]
\label{H_K} \\
H_P & := &  -\int_{S^1} \mbox{d} \theta N (\theta) \f{1}{\sqrt{E}}
\left[ -\f{1}{4} \left( \partial_\theta {\cal E} \right)^2 + \f{({\cal
E})^2}{4} \left(\f{\partial_\theta E^x}{ E^x} - \f{\partial_\theta E^y}{
E^y}\right)^2 \right] \nonumber \\
& &  \hspace{0.0cm} -\int_{S^1} \mbox{d} \theta N (\theta)
\f{1}{2}\del_\theta\left[\f{2 {\cal E} \left(\del_\theta {\cal E}
\right) }{\sqrt{E}}\right] \label{H_P}
\end{eqnarray} 
In the above we have used the expressions of $\Gamma_x$ and $\Gamma_y$
and suppressed the terms dependent on the Gauss constraint which will
drop out on gauge invariant basis states. Only $H_K$ depends on the
configuration variables and all terms have two powers of momenta in the
numerator and the $\sqrt{E}$ in the denominator whose vanishing is a
potential problem.

The kinetic term has a structure similar to the Euclidean term in the
full theory $\sim E E F/\sqrt{q}$ (but it is {\em not} the
simplification of the Euclidean term of LQG). This will be treated in a
manner similar to the full theory, using appropriate holonomies in the
form $h_ih_jh^{-1}_i h_j^{-1} h_k\{h_k^{-1}, V\}$. The remaining terms
are functions of momenta alone and the $E^{-1/2}$ is treated using
Poisson bracket of the volume with suitable holonomy, 

Although the holonomies defined before, eg in the basis states
(\ref{BasisStates}), are all phases (Abelian gauge theory), it is
convenient to introduce their $SU(2)$ valued analogues using the $\eta$
dependent $\tau$ matrices defined in eq.  (\ref{TauDef}) and in eq. 
(\ref{tauxtauy}). Thus,
\begin{eqnarray}
h_\theta (\mathcal{I}) & := & \exp \left(\tau_3 \int_{\mathcal{I}} {\cal
A} (\theta') ~{\mbox d} \theta' \right) ~ = ~
\mathrm{cos}\left(\f{1}{2}\int_{\mathcal{I}} \mathcal{A}\right) + 2
\tau_3 \ \mathrm{sin}\left(\f{1}{2}\int_{\mathcal{I}} \mathcal{A}\right)
\nonumber \\ 
h_x(\theta) & := & \exp \left( \mu_0 X(\theta) \ \tau_x (\theta) \right) ~ = ~
\mathrm{cos}\left(\f{\mu_0}{2} X(\theta)\right) + 2\tau_x(\theta) \
\mathrm{sin}\left(\f{\mu_0}{2} X(\theta)\right) \nonumber \\
h_y(\theta) & := & \exp \left( \nu_0 Y(\theta) \ \tau_y (\theta)  \right) ~ = ~
\mathrm{cos}\left(\f{\nu_0}{2} Y(\theta)\right) + 2\tau_y (\theta) \
\mathrm{sin}\left(\f{\nu_0}{2} Y(\theta)\right) \label{holonomies}
\end{eqnarray}
Each of the sin, cos are well defined on the kinematical Hilbert space
(this was the reason for the factors of $1/2$ in the definitions of the
holonomies in the basis states) and therefore so are the above
$SU(2)$-valued holonomies.  The interval $\mathcal{I}$ will typically be
a cell of a partition, $(\theta_i, \theta_i + \epsilon)$.  The
parameters $\mu_0, \nu_0$ are the chosen and fixed representations of
$\mathbb{R}_{\mathrm{Bohr}}, k_0 = 1$ is the fixed representation of the
$U(1)$, while $\epsilon $ is a small parameter which will also play the
role of the regulator parameter.  Let us also define, the volume
function labelled by an interval $\mathcal{I}$ and a point $\theta$ inside 
the interval:
\begin{equation}
V(\mathcal{I},\theta) := \sqrt{|\mathcal{E}(\theta)| \left|\int_{\mathcal{I}}
E^x\right| \left|\int_{\mathcal{I}} E^y\right| } 
\end{equation}
For brevity of notation we will suppress the label $\theta$ and denote the 
above volume function as $V(\mathcal{I})$.

Consider expression of the form Tr($h_ih_jh_i^{-1}h_j^{-1}h_k\{h_k^{-1},
\sqrt{E}\}$) for distinct $i, j, k$ taking values $\theta, x, y$.  For
small values of $X, Y, \int_{\mathcal{I}}\mathcal{A}$, the holonomies
can be expanded in a power series. Because of the trace, it is enough to
expand each holonomy up to 1st order. The surviving terms are quadratic
terms arising from products of the linear ones and a linear term coming
from $h_k$. If one interchanges the $i \leftrightarrow j $ holonomies,
the linear term retains the sign while the quadratic one changes the
sign. Thus taking the difference of the two traces, leaves us with only
the quadratic terms which are exactly of the form needed in $H_K$.
Explicitly,
\begin{eqnarray}
\mathrm{Tr} \left[ \left\{~ h_x(\theta) h_y(\theta) h_x^{-1}(\theta) h_y^{-1}(\theta) -
h_y(\theta) h_x(\theta) h_y^{-1}(\theta) h_x^{-1}(\theta)~ \right\}
h_{\theta}(\mathcal{I})\{h_{\theta}^{-1}(\mathcal{I}), V(\mathcal{I})\}
\right] \nonumber
\end{eqnarray}
\begin{eqnarray} \label{XY-plaquette}
\hspace{3.0cm} \approx ~ \left(-\f{\kappa'\gamma}{2} \
\mu_0\nu_0\right)  \ \f{\epsilon \ X(\theta)Y(\theta) E^x(\theta)
E^y(\theta)}{\sqrt{E(\theta)}} 
\end{eqnarray}
\begin{eqnarray}
\mathrm{Tr} \left[ \left\{~ h_y(\theta) h_{\theta}(\mathcal{I}) h_y^{-1}(\theta +
\epsilon) h_{\theta}^{-1}(\mathcal{I}) - h_{\theta}(\mathcal{I}) h_y( \theta
+ \epsilon) h_{\theta}^{-1}(\mathcal{I}) h_y^{-1}(\theta) ~ \right\}
h_{x}(\theta)\{h_{x}^{-1}(\theta), V(\mathcal{I})\} \right] \nonumber
\end{eqnarray}
\begin{eqnarray} \label{YTheta-plaquette}
\hspace{3.0cm} \approx ~ \left(- \f{\kappa'\gamma}{2}
\mu_0\nu_0\right) \ \f{\epsilon \ Y(\theta)(\mathcal{A}(\theta) +
\partial_{\theta}\eta(\theta) ) E^y(\theta) \mathcal{E}(\theta)}{\sqrt{E(\theta)}} 
\end{eqnarray}
\begin{eqnarray}
\mathrm{Tr} \left[ \left\{~ h_{\theta}(\mathcal{I}) h_x(\theta + \epsilon)
h_{\theta}^{-1}(\mathcal{I}) h_x^{-1}(\theta) - h_x(\theta)
h_{\theta}(\mathcal{I}) h_x^{-1}(\theta + \epsilon)
h_{\theta}^{-1}(\mathcal{I}) ~\right\} h_{y}(\theta)\{h_{y}^{-1}(\theta),
V(\mathcal{I})\} \right]\nonumber 
\end{eqnarray}
\begin{eqnarray} \label{XTheta-plaquette}
\hspace{3.0cm} \approx ~ \left(- \f{\kappa'\gamma}{2}
\mu_0\nu_0\right) \ \f{\epsilon \ (\mathcal{A}(\theta) +
\partial_{\theta}\eta(\theta)) X(\theta) \mathcal{E}(\theta) 
E^x(\theta)}{\sqrt{E(\theta)}}
\end{eqnarray}
In equations (\ref{YTheta-plaquette}) and (\ref{XTheta-plaquette}) 
${\mathcal I}$ is the interval between $\theta$ and $\theta + \epsilon$. 
The derivatives of $\eta$ arise from the position dependence of the
$\tau_x, \tau_y$ matrices which satisfy:
\begin{eqnarray}
\tau_x(\theta + \epsilon) - \tau_x(\theta) & \approx & \epsilon\del_{\theta}\tau_x
~ = ~ \epsilon\partial_{\theta}\eta ~ \tau_y(\theta) \nonumber \\
\tau_y(\theta + \epsilon) - \tau_y(\theta) & \approx & \epsilon\del_{\theta}\tau_y
~ = ~ - \epsilon\partial_{\theta}\eta ~ \tau_x(\theta) \label{TauDerivative}
\end{eqnarray}
In the above, we have also used:
\begin{eqnarray}
h_x(\theta)\{h_x(\theta)^{-1}, V(\mathcal{I})\} & = &
-\f{\kappa'\gamma}{2}\mu_0\tau_x \f{ \mathcal{E}(\theta)
\int_{\mathcal{I}}E^y }{V(\mathcal{I})} ~ \approx ~
-\f{\kappa'\gamma}{2}\mu_0\tau_x \f{E^y(\theta)
\mathcal{E}(\theta)}{\sqrt{E(\theta)}} \nonumber \\
h_y(\theta)\{h_y(\theta)^{-1}, V(\mathcal{I})\} & = &
-\f{\kappa'\gamma}{2}\nu_0\tau_y \f{ \mathcal{E}(\theta)
\int_{\mathcal{I}}E^x }{V(\mathcal{I})} ~ \approx ~
-\f{\kappa'\gamma}{2}\nu_0\tau_y \f{E^x(\theta)
\mathcal{E}(\theta)}{\sqrt{E(\theta)}} \nonumber \\
h_{\theta}\{h_{\theta}^{-1}, V(\mathcal{I})\} & = &
-\f{\kappa'\gamma}{2}\tau_3 \f{\int_{\mathcal{I}}E^x
\int_{\mathcal{I}}E^y}{V(\mathcal{I})} ~   ~ \approx ~
-\f{\kappa'\gamma}{2}\epsilon\tau_3 \f{E^x(\theta) E^y(\theta)}
{\sqrt{E(\theta)}} \label{ApproxOne}\\
\int_{\mathcal{I}}\mathcal{A} ~ \approx ~ \epsilon \mathcal{A}(\theta) &,&
\int_{\mathcal{I}}E^x ~ \approx ~ \epsilon E^x(\theta) \hspace{0.7cm} ~,~ ~
\int_{\mathcal{I}}E^y ~ \approx ~ \epsilon E^y(\theta)\label{ApproxTwo} 
\end{eqnarray}

In the quantization of the $H_P, H_T$, we also use the following
identities repeatedly (in the form LHS/RHS = 1):
\begin{eqnarray} 
\mathcal{Z}(\mathcal{I}) & := & \epsilon^{abc}\mathrm{Tr}\left[~
h_a\{h_a^{-1}, V(\mathcal{I})\}~ h_b\{h_b^{-1}, V(\mathcal{I})\}~
h_c\{h_c^{-1}, V(\mathcal{I})\}~ \right] \nonumber \\
& = & \f{3}{2} \left(\f{\kappa'\gamma}{2}\right)^3 \mu_0\nu_0
V(\mathcal{I}) \label{Identity1} \\
\mathcal{Z}_{\alpha}(\mathcal{I}) & := &
\epsilon^{abc}\mathrm{Tr}\left[~ h_a\{h_a^{-1},
({V(\mathcal{I}))^{\alpha}}\}~ h_b\{h_b^{-1},
({V(\mathcal{I}))^{\alpha}}\}~ h_c\{h_c^{-1},
({V(\mathcal{I}))^{\alpha}}\}~ \right] \nonumber \\ 
& = & \f{3}{2} \left(\f{\kappa'\gamma}{2}\right)^3 \mu_0\nu_0 \
\alpha^3 (V(\mathcal{I}))^{3 \alpha - 2} \nonumber \\
& = & \alpha^3 (V(\mathcal{I}))^{3(\alpha - 1)}~
\mathcal{Z}(\mathcal{I}) \label{Identity2}
\end{eqnarray}

These are essentially versions of the identity $ 1 =
\left(\Case{|\det(e_a^i)|}{\sqrt{E}}\right)^n$ \cite{QSD5}.  Having
noted the ingredients common to the quantization of the different pieces
of the Hamiltonian constraint, we turn to each one in explicit details.

\subsection{Quantization of $H_K$}

Choosing a partition of $S^1$ with a sufficiently large number of $n$
points at $\theta_i, i = 1, \cdots, n, \theta_n = 2\pi, \ \epsilon =
\theta_{i + 1} - \theta_i $, we write the integral as a sum,
\begin{eqnarray}
H_K & \approx & 
\f{1}{\gamma^2} \sum_{i = 1}^n  \epsilon N (\bar \theta_i)
\f{1}{\sqrt{E}(\bar \theta_i)} \left[  X E^x Y E^y + \left({\cal A} +
\partial_{\theta}\eta\right) {\cal E}( X E^x  + Y E^y) \right](\bar
\theta_i) \label{H_KSum} \\
& = & 
\f{1}{\gamma^2} \sum_{i = 1}^n  N (\bar \theta_i)
\f{1}{\sqrt{\epsilon^2 E}(\bar \theta_i)} \left[  X (\epsilon E^x) Y
(\epsilon E^y) + \epsilon \left({\cal A} +
\partial_{\theta}\eta\right) {\cal E}( X (\epsilon E^x)  + Y
(\epsilon E^y)) \right] \nonumber \\
& = & 
\f{1}{\gamma^2} \sum_{i = 1}^n N (\bar \theta_i) \f{1}{V(\mathcal{I}_i)}
\left[  X(\bar \theta_i) \left(\int_{\mathcal{I}_i}E^x\right) Y(\bar
\theta_i) \left( \int_{\mathcal{I}_i}E^y\right)  + \right. \nonumber \\
& & 
\left. \hspace{3.5cm} \left(\int_{\mathcal{I}_i}{\cal A} +
\partial_{\theta}\eta\right) {\cal E}(\bar \theta_i) \left\{ X(\bar
\theta_i) \int_{\mathcal{I}_i}E^x  + Y(\bar \theta_i)
\int_{\mathcal{I}_i}E^y \right\} \right]
\end{eqnarray} 

From the equations (\ref{XY-plaquette}, \ref{YTheta-plaquette},
\ref{XTheta-plaquette}), one sees immediately that for small values of
the extrinsic curvature components ($ \sim X, Y$, classical regime) and
sufficiently refined partition ($\epsilon \ll 1$, continuum limit), the
$i^{\mathrm{th}}$ term in the sum can be written in terms of the traces
of the $SU(2)$ valued holonomies. In other words, the expression in
terms of holonomies and fluxes, does go over to the classical expression
in the classical regime and can be promoted to an operator by putting
hats on the holonomies and fluxes and replacing Poisson brackets by
$(i\hbar)^{-1}$ times the commutators.  Here, there are possibilities
for choosing the ordering of various factors and in this paper, we will
make the ``standard choice'' of putting the holonomies on the left.
Thus, the regulated quantum operator corresponding to the kinetic piece
is (suppressing the hats on the holonomies and using $\lP^2 :=
\kappa'\hbar$),
\begin{eqnarray}
\widehat H_K^{\mathrm{reg}} & := & 
i\f{2}{\lP^2\gamma^3}\f{1}{\mu_0\nu_0} \sum_{i = 1}^n N (\bar \theta_i)
\mathrm{Tr}\bigg(~  \nonumber \\
& & \hspace{0.4cm} \left\{h_x h_y h_x^{-1} h_y^{-1} - h_y h_x h_y^{-1}
h_x^{-1}\right\} h_{\theta}(\mathcal{I}_i) \left[
h_{\theta}^{-1}(\mathcal{I}_i), \hat V(\mathcal{I}_i) \right] \\ 
& & + \left\{h_y h_{\theta}(\mathcal{I}_i)h_y^{-1}(\bar \theta_{i} +
\epsilon) h_{\theta}^{-1}(\mathcal{I}_i) -
h_{\theta_i}(\mathcal{I}_i)h_y(\bar \theta_{i} + \epsilon)
h_{\theta}^{-1}(\mathcal{I}_i)h_y^{-1}\right\} h_x \left[ h_x^{-1} ,
\hat V(\mathcal{I}_i) \right] \nonumber \\
& & + \left\{h_{\theta}(\mathcal{I}_i)h_x(\bar \theta_i +
\epsilon)h_{\theta}^{-1}(\mathcal{I}_i) h_x^{-1}  -  h_x
h_{\theta}(\mathcal{I}_i)h_x^{-1}(\bar \theta_i +
\epsilon)h_{\theta}^{-1}(\mathcal{I}_i)\right\} h_y \left[ h_y^{-1},
\hat V(\mathcal{I}_i) \right] \bigg)\nonumber
\end{eqnarray} 
In the above equation, the point holonomies without an explicit
argument, are at $\bar \theta_i$. 

At this point it is convenient to define the following families of
operators:
\begin{eqnarray}
\hat {\cal O}_{\alpha}^x({\cal I},\theta) & := & \left[ \cos \left(
\f{1}{2} \mu_0 X(\theta)\right) \hat V^{\alpha}({\cal I}) \sin \left
(\f{1}{2} \mu_0 X(\theta)\right) - \right. \nonumber \\
& & 
\left. \hspace{4.0cm} \sin\left (\f{1}{2} \mu_0 X(\theta) \right) \hat
V^{\alpha}({\cal I}) \cos \left( \f{1}{2} \mu_0 X(\theta) \right)
\right] \nonumber \\
\hat {\cal O}_{\alpha}^y({\cal I},\theta) & := & \left[ \cos \left(
\f{1}{2} \mu_0 Y(\theta)\right) \hat V^{\alpha}({\cal I}) \sin \left
(\f{1}{2} \mu_0 Y(\theta)\right) - \right. \nonumber \\
& & 
\left. \hspace{4.0cm} \sin\left (\f{1}{2} \mu_0 Y(\theta) \right) \hat
V^{\alpha}({\cal I}) \cos \left( \f{1}{2} \mu_0 Y(\theta) \right)
\right] \nonumber \\
\hat {\cal O}_{\alpha}^{\theta}({\cal I},\theta) & := & \left[ \cos
\left( \f{1}{2} \int_{ {\cal I} }{\cal A} \right) \hat V^{\alpha}({\cal
I}) \sin \left (\f{1}{2} \int_{ {\cal I} }{\cal A} \right) - \right.
\nonumber \\
& & 
\left. \hspace{4.0cm} \sin\left (\f{1}{2} \int_{ {\cal I} }{\cal A}
\right) \hat V^{\alpha}({\cal I}) \cos \left( \f{1}{2} \int_{ {\cal I}
}{\cal A} \right) \right] 
\end{eqnarray}
In the above, $\theta$ is a point in the interval ${\cal I}$ and $\alpha
> 0$ is the power of the volume operator. Again for simplicity of
notation we will suppress the $\theta$ labels in the above operators. 

The operator form of ${\cal Z}_{\alpha}({\cal I})$ can be obtained as:
\begin{eqnarray}
\hat{\cal Z}_{\alpha}({\cal I}) & := &  \epsilon^{abc} \mbox{Tr}\bigg(\
\hat{h}_a\ [\ \hat{h}_a^{-1}\ ,\ {\hat{V}(\cal I)}^{\alpha}\ ]\
\hat{h}_b\ [\ \hat{h}_b^{-1}\ ,\ {\hat{V}(\cal I)}^{\alpha}\ ] \
\hat{h}_c\ [\ \hat{h}_c^{-1}\ ,\ {\hat{V}(\cal I)}^{\alpha}\ ]\  \bigg)
\\
& = &  -12 ~ \hat {\cal O}_{\alpha}^x({\cal I})\ \hat {\cal
O}_{\alpha}^y({\cal I}) \ \hat {\cal O}_{\alpha}^{\theta}({\cal I})
\end{eqnarray}

Using the expressions for the holonomies in terms of the
``trigonometric'' operators given in the eq. (\ref{holonomies}),  and
evaluating the traces etc, one can see that,
\begin{eqnarray}
\widehat H_K^{\mathrm{reg}} & = & 
- i\f{4}{\lP^2\gamma^3}\f{1}{\mu_0\nu_0} \sum_{i = 1}^n N (\bar
\theta_i) \Bigg[ \left\{\ \sin\left(\mu_0 X(\bar \theta_i) \right)\sin
\left(\nu_0 Y(\bar \theta_i) \right)\ \right\} \times {\cal
O}_1^{\theta}({\cal I}_i) ~ + ~
\nonumber \label{1stbracket} \\
& & \hspace{1.0cm}\left\{2 \sin \left (\f{1}{2} \nu_0 Y(\bar \theta_i +
\epsilon) \right)\cos \left(\f{1}{2} \nu_0 Y(\bar \theta_{i})\right)\sin
\left( \int_{\mathcal{I}_i} {\cal A} -\Delta_i\right)\right\}\times
{\cal O}_1^{x}({\cal I}_i) ~ + ~
\nonumber \label{2ndbracket}\\
& & \left. \hspace{1.0cm}\left\{2 \sin \left (\f{1}{2} \mu_0 X (\bar
\theta_i + \epsilon) \right)\cos \left(\f{1}{2} \mu_0 X(\bar \theta_{i
})\right)\sin \left( \int_{\mathcal{I}_i} {\cal A} -\Delta_i\right)
\right\} \times {\cal O}_1^{y}({\cal I}_i) \right]
\nonumber \label{3rdbracket}
\end{eqnarray} 
In the above $\Delta_i := \eta(\bar \theta_i) - \eta(\bar \theta_i +
\epsilon)$ and is outside the integral.

\subsection{Quantization of $H_P$}

All the three terms of $H_P$ are functions of the momenta (triad) only.
These have to be expressed in terms of fluxes and holonomies alone.
Furthermore, the power(s) of momenta in the denominators will make the
action on some states to be singular. The first part is easy to take
care of thanks to the density weight 1. For the second part we use the
by now familiar procedure of using the identities (\ref{Identity1},
\ref{Identity2}).  Due to the spatial dimension being 1, it is easier to
convert triads in terms of fluxes directly, without explicitly doing any
point-splitting (one could of course do that if so desired \cite{QSD5}).

The terms in the $H_P$ will be manipulated in the following steps: (i)
introduce sufficient number, $k > 0$, of positive powers of $ 1 = 16( 3
(\kappa'\gamma)^3 \mu_0\nu_0)^{-1}
\mathcal{Z}(\mathcal{I})/V(\mathcal{I})$ and express $\mathcal{Z}$ in
terms of $\mathcal{Z}_{\alpha}$. This introduces further powers of the
volume; (ii) choose $\alpha(k)$ such that explicit multiplicative
factors of the volume become 1 and further choose $k$ sufficiently large
so that $\alpha(k) > 0$ is obtained. Note that the choice of $k > 0$
constitutes a quantization ambiguity. Now the expression can be promoted
to an operator. Here are the details.
\\ \\ {\underline {\it The first term of $H_P$}~:}

\begin{eqnarray}
-\int_{S^1} N (\theta) \f{1}{\sqrt{E(\theta)}} \left[ -\f{1}{4} \left(
\partial_\theta {\cal E} \right)^2 \right] & \approx & +\f{1}{4} \sum_{i
= 1}^n N(\bar \theta_i) \epsilon \f{ \left(
\del_{\theta}\mathcal{E}(\bar \theta_i)\right)^2}{\sqrt{E(\bar
\theta_i)}}
~ = ~ \f{1}{4} \sum_{i = 1}^n N(\bar \theta_i) \f{
\left(\epsilon\del_{\theta}\mathcal{E}(\bar \theta_i)\right)^2}{\epsilon
\sqrt{E(\bar \theta_i)}} \nonumber \\
& = & \f{1}{4} \sum_{i = 1}^n N(\bar \theta_i) \f{ \left(
\mathcal{E}(\bar \theta_i + \epsilon) - \mathcal{E}(\bar \theta_i)
\right)^2}{\sqrt{\mathcal{E}(\bar \theta_i) \ \int_{ {\cal I}_i} E^x \
\int_{ {\cal I}_i} E^y }} ~ (1)^k
\end{eqnarray}
\begin{eqnarray} \label{FirstTerm}
\mathrm {RHS} & = & \f{1}{4} \sum_{i = 1}^n N(\bar \theta_i) \f{ \left(
\mathcal{E}(\bar \theta_i + \epsilon) - \mathcal{E}(\bar \theta_i)
\right)^2}{V(\mathcal{I}_i)} \left(\f{16}{3
(\kappa'\gamma)^3\mu_0\nu_0}\right)^k
\left(\f{\mathcal{Z}(\mathcal{I}_i)}{V(\mathcal{I}_i)}\right)^k \\
& = & \f{1}{4} \left(\f{16}{3 (\kappa'\gamma)^3\mu_0\nu_0}\right)^k
\sum_{i = 1}^n N(\bar \theta_i) \f{ \left( \mathcal{E}(\bar \theta_i +
\epsilon) - \mathcal{E}(\bar \theta_i) \right)^2}{V(\mathcal{I}_i)}
\left(\f{\mathcal{Z}_{\alpha}(\mathcal{I}_i)}{\alpha^3
(V(\mathcal{I}_i))^{(3\alpha - 2)} }\right)^k \nonumber \\
& = & \left. \f{1}{4} \left(\f{16}{3 (\kappa'\gamma)^3\mu_0\nu_0
\alpha^3}\right)^k \sum_{i = 1}^n N(\bar \theta_i) \left(
\mathcal{E}(\bar \theta_i + \epsilon) - \mathcal{E}(\bar \theta_i)
\right)^2 \left(\mathcal{Z}_{\alpha}(\mathcal{I}_i)\right)^k
\right|_{\alpha := \f{2}{3} - \f{1}{3 k}} \nonumber
\end{eqnarray}
In the last line we have chosen $\alpha := \Case{2}{3} - \Case{1}{3 k}$
which removes the explicit factors of the volume. The choice of $k > 0$
is limited by $\alpha > 0$ (being a power of the volume appearing in
$\mathcal{Z}_{\alpha}$). Some convenient choices would be $k = 1 \
(\alpha = 1/3), \ k = 2 \ (\alpha = 1/2)$ etc. For all such choices, the
above expression can be promoted to a well defined operator. 
\\ \\ {\underline {\it The second term of $H_P$}~:} \\

To begin with one observes that $E^y/E^x$ is a scalar,
$\del_{\theta}\ln(E^y/E^x)$ is a scalar density. This term is then
manipulated as:
\begin{eqnarray}
- \f{1}{4} \int_{S^1} N (\theta) \f{({\cal
E}(\theta))^2}{\sqrt{E(\theta)}} \left( \f{\partial_\theta E^x}{E^x} -
\f{\del_{\theta} E^y}{E^y} \right)^2 
& = & - \f{1}{4} \int_{S^1} N (\theta) \f{({\cal
E}(\theta))^2}{\sqrt{E(\theta)}} \left(\partial_\theta \ln
\left(\f{E^y}{E^x}\right)\right)^2 
\end{eqnarray}
\begin{eqnarray} \label{SecondTerm}
\mathrm{RHS} & \approx & - \f{1}{4} \sum_{i = 1}^n N(\bar \theta_i)
\epsilon \f{({\cal E}(\bar \theta_i))^2}{\sqrt{E(\bar \theta_i)}}
\left(\partial_\theta \ln \left(\f{E^y}{E^x} (\bar \theta_i)
\right)\right)^2 \nonumber \\
& = & - \f{1}{4} \sum_{i = 1}^n N(\bar \theta_i) \f{({\cal E}(\bar
\theta_i))^2}{\sqrt{\epsilon^2 E(\bar \theta_i)}} \left( \f{E^x(\bar
\theta_i)}{E^y(\bar \theta_i)}\ \epsilon \partial_\theta
\left(\f{E^y}{E^x} (\bar \theta_i) \right)\right)^2 \nonumber \\
& = & - \f{1}{4} \sum_{i = 1}^n N(\bar \theta_i) \f{({\cal E}(\bar
\theta_i))^2}{V(\mathcal{I}_i)} \left[ \f{E^x(\bar \theta_i)}{E^y(\bar
\theta_i)}\ \Bigg\{ \left.\f{E^y}{E^x}\right|_{\bar \theta_i + \epsilon}  -
\left.\f{E^y}{E^x}\right|_{\bar \theta_{i}} \Bigg\} \right]^2 \nonumber
\\
& = & - \f{1}{4} \sum_{i = 1}^n N(\bar \theta_i) \f{({\cal E}(\bar
\theta_i))^2}{V(\mathcal{I}_i)} \left[ \f{\int_{ {\cal I}_i} E^x}{\int_{
{\cal I}_i} E^y}\ \Bigg\{ \f{\int_{ {\cal I}_{i + 1} } E^y}{\int_{ {\cal
I}_{i + 1} } E^x}  - \f{\int_{ {\cal I}_{i} } E^y}{\int_{ {\cal I}_{i} }
E^x} \Bigg\} \right]^2
\end{eqnarray}
Now we have the fluxes in the denominator which can be defined exactly
as the inverse triad operators of LQC \cite{InverseTriad}. To be
explicit, denoting the fluxes as ${\cal F}_{x,{\cal I}} := \int_{\cal I}
E^x, {\cal F}_{y,{\cal I}} := \int_{\cal I} E^y$.
\begin{eqnarray}
{\cal F}^{-1}_{x,{\cal I}} & = & \left(\f{1}{\kappa'\gamma l}
\right)^{\f{1}{1 - l}} \left\{ X(v),  {\cal F}^l_{x,{\cal I}}
\right\}^{\f{1}{1 - l}} ~ ~, \hspace{2.0cm} l \in (0, 1) \nonumber \\
& = & \left(\f{2 i}{\kappa'\gamma l \mu_0} \right)^{\f{1}{1 - l}} \Bigg(
h_v^{(\mu_0/2)}(X)\left\{ h_v^{(- \mu_0/2)}(X), {\cal F}^l_{x,{\cal I}}
\right\}\Bigg)^{\f{1}{1 - l}} 
\end{eqnarray}
and similarly for ${\cal F}^{-1}_{y,{\cal I}}$. These can be promoted to
a well defined operator. Continuing with the equation above, 
\begin{eqnarray} \label{SecondTermDetailed}
\mathrm{RHS} & = & - \f{1}{4} \sum_{i = 1}^n N(\bar \theta_i) \f{({\cal
E}(\bar \theta_i))^2}{V(\mathcal{I}_i)} 
\left[ {\cal F}^{-1}_{y,{\cal I}_i} {\cal F}_{x,{\cal I}_i} \Bigg( {\cal
F}^{-1}_{x,{ {\cal I}_{i + 1}}} {\cal F}_{y,{\cal I}_{i + 1}}  - {\cal
F}^{-1}_{x,{\cal I}_{i}} {\cal F}_{y,{\cal I}_{i}} \Bigg) \right]^2
\nonumber \\
& = & - \f{1}{4} \left(\f{16}{3 (\kappa'\gamma)^3\mu_0\nu_0
\alpha^3}\right)^k\  \sum_{i = 1}^n N(\bar \theta_i) ({\cal E}(\bar
\theta_i))^2 \times \\ 
& & \hspace{2.0cm} 
\left[ {\cal F}^{-1}_{y,{\cal I}_i} {\cal F}_{x,{\cal I}_i} \Bigg( {\cal
F}^{-1}_{x,{ {\cal I}_{i + 1}}} {\cal F}_{y,{\cal I}_{i + 1}}  - {\cal
F}^{-1}_{x,{\cal I}_{i}} {\cal F}_{y,{\cal I}_{i}} \Bigg) \right]^2
\left.  \left(\mathcal{Z}_{\alpha}(\mathcal{I}_i) \right)^k
\right|_{\alpha = \f{2}{3} - \f{1}{3 k}}  \nonumber 
\end{eqnarray}
where, in the last step, we have manipulated,
\begin{eqnarray}
\f{1}{V(\mathcal{I}_i)} ~ = ~ \f{1}{V(\mathcal{I}_i)} (1)^k & = &
\f{1}{V(\mathcal{I}_i)} \left(\f{16}{3
(\kappa'\gamma)^3\mu_0\nu_0}\right)^k
\left(\f{\mathcal{Z}(\mathcal{I}_i)}{V(\mathcal{I}_i)}\right)^k
\nonumber \\
& = & \left.  \left(\f{16}{3 (\kappa'\gamma)^3\mu_0\nu_0
\alpha^3}\right)^k\  \left(\mathcal{Z}_{\alpha}(\mathcal{I}_i) \right)^k
\right|_{\alpha = \f{2}{3} - \f{1}{3 k}}  
\end{eqnarray}
The choice of $\alpha$ would be same as that in the first term.  \\ \\ 
{\underline {\em The third term of $H_P$}~:}

\begin{eqnarray}
H_T ~ = ~ - \int_{S^1} N(\theta)
\del_\theta\left[\f{\mathcal{E}\del_{\theta}\mathcal{E}}{\sqrt{E(\theta)}}\right]
& \approx & - \sum_{i = 1}^n N(\bar \theta_i) \epsilon
\del_{\theta}\left[\f{\mathcal{E}(\bar
\theta_i)\del_{\theta}\mathcal{E}}{\sqrt{E(\bar \theta_i)}}\right]
\nonumber 
\end{eqnarray}
\begin{eqnarray} \label{HD}
\mathrm{RHS} & = & - \sum_{i = 1}^n N(\bar \theta_i) \left[ \left.
\left\{\f{\mathcal{E}\epsilon\del_{\theta}\mathcal{E}}{\epsilon
\sqrt{E}}\right\}\right|_{\bar \theta_{i} + \epsilon} -  \left.
\left\{\f{\mathcal{E}\epsilon\del_{\theta}\mathcal{E}}{\epsilon
\sqrt{E}}\right\}\right|_{\bar \theta_{i}} \right] \nonumber \\
& = & - \sum_{i = 1}^n N(\bar \theta_i) \left[ \f{\mathcal{E}(\bar
\theta_i + \epsilon) \left\{ \mathcal{E}(\bar \theta_{i} + 2\epsilon ) -
\mathcal{E}(\bar \theta_i + \epsilon )~ \right\} }{ V(\mathcal{I}_{i +
1})} -  \f{\mathcal{E}(\bar \theta_{i}) \left\{ \mathcal{E}(\bar
\theta_i + \epsilon) - \mathcal{E}(\bar \theta_{i})~ \right\} }{
V(\mathcal{I}_{i})} \right] \nonumber \\
& = & - \left(\f{16}{3 (\kappa'\gamma)^3\mu_0\nu_0 \alpha^3}\right)^k\
\sum_{i = 1}^n N(\bar \theta_i) \left[ {\mathcal{E}(\bar \theta_i +
\epsilon) \left\{ \mathcal{E}(\bar \theta_i + 2\epsilon) -
\mathcal{E}(\bar \theta_i + \epsilon) \right\} }(\
\mathcal{Z}_{\alpha}(\mathcal{I}_{i + 1})\ )^{k} \right. \nonumber \\
& & \hspace{4.5cm} - \left. \left. \mathcal{E}(\bar \theta_{i}) \left\{
\mathcal{E}(\bar \theta_i + \epsilon) - \mathcal{E}(\bar \theta_{i})~
\right\} (\ {\cal Z}_{\alpha}(\mathcal{I}_{i})\ )^k \right]
\right|_{\alpha = \f{2}{3} - \f{1}{3 k}} 
\end{eqnarray}

At this point we have expressed the $H_P$ in terms of the holonomy-flux
variables and quantization can be carried out simply by replacing the
$({\cal Z}_{\alpha})^{k} \to (-i/\hbar)^{3k} (\hat {\cal
Z}_{\alpha})^{k}$. This correctly combines with the powers of $\kappa'$
to give $(\lP^2)^{3k}$ in the denominator. The $({\cal
Z}_{\alpha})^{k}$, will give $(\lP^{3\alpha})^{3k}$ since each factor of
volume gives $3 \alpha$, there are 3 factors of volume in each ${\cal
Z}$ and there is the overall power of $k$. Substituting for $\alpha$ one
sees that each of the terms in $H_P, H_T$ has $\lP^{-3}$ apart from the
$\lP^4$ supplied by the factors of momenta/fluxes, thus giving the
correct dimensions.

The operators ${\cal O}_{\alpha}^{a} := [\cos(\cdots) \hat V^{\alpha}
\sin(\cdots) - \sin(\cdots) \hat V^{\alpha} \cos(\cdots)],\ a = x, y,
\theta, $ appear in all the terms and is a function of both holonomies
and fluxes. To see that this is actually diagonal in the charge network
basis, write the cos and sin operators as sums and differences of the
exponentials (i.e. holonomies).  It then follows that,
\begin{eqnarray}
\cos(\cdots) \hat V^{\alpha} \sin(\cdots) - \sin(\cdots) \hat V^{\alpha}
\cos(\cdots)] & = & \f{1}{2i}\left[ e^{-i(\cdots)} \hat V e^{+i(\cdots)}
- e^{+i(\cdots)} \hat V e^{-i(\cdots)} \right]
\end{eqnarray}
It is now obvious that the operators are diagonal and thus commute with
all the flux operators.  Thus there are no ordering issues in
quantization of $H_P$ operators. In the $H_K$, however, operators of the
above type are ordered on the right as in LQG. 

\section{Action on States}

To make explicit the action of the Hamiltonian constraint on the basis
states, it is useful to make a couple of observations. Every
gauge-invariant {\em basis} state can be thought of as a collection of
$m-$vertices with a quadruple of labels ($k_v^{\pm}, \mu_v, \nu_v$), all
non-zero. The $k_v^{\pm}$ denoting the U(1) representations on the two
edges meeting at $v$ with $+$ referring to the exiting edge and $-$ to
the entering edge. A partition may also be viewed as a graph except that
at its ``vertices'' all representation labels are zero. Secondly, the
action of the flux operators labelled by ${\cal I}$, on a basis state is
necessarily zero if none of the vertices of the state have an
intersection with the label interval. Note that the operator ${\cal
E}(\theta_i)$, however always has a non-zero action on a basis state.
This is because, all graphs are closed and hence there is always an edge
(and non-zero label for a basis state) which overlaps with $\theta_i$.
The volume operator associated with an interval ${\cal I}$ gives a
non-zero contribution on a basis state {\em only if} ${\cal I}$ contains
a vertex of the graph. Recall that our partition is sufficiently refined
so that each cell contains {\em at most one vertex} (two vertices at the
cell boundaries are counted as a single vertex in the interior).

The full Hamiltonian has been written as a sum using a partition of
$S^1$. Consider the $i^{\rm th}$ term in each of the $H_K, H_P$.  Each
of these contains ${\cal O}_{\alpha}^a$ operators either separately (as
in $H_K$) or as a product through the ${\cal Z}_{\alpha}$ (as in $H_P$).
Since these contain the volume operator, it ensures that the action of
each of these terms is necessarily zero unless the ${\cal I}_i$ contains
a vertex of the basis state. Evidently the action of the full constraint
is {\em finite} regardless of the chosen partition. 

The factors of trigonometric operators multiplying the ${\cal
O}_{\alpha}^a$ on the left in $H_K$, can be thought of as ``creating new
vertices'' at the points $\bar \theta_i$ of the partition. Notice
however that at these new vertices one has either an edge holonomy or
{\em one} of the point holonomies only i.e. the volume operator acting
at these vertices will give zero.

Summarizing, thanks to the $\hat {\cal O}, \hat {\cal Z}$ operators
acting first, only those intervals of a partition will contribute which
contain at least one vertex of the graph of a basis state. This
immediately implies that in the second term of $H_P$ (eq.
\ref{SecondTermDetailed}), only one of the terms in square bracket will
contribute. We will return to this later.  Let us denote the factor
associated with a vertex $v$ of a basis state by $|k^{\pm}_v, \mu_v,
\nu_v\rangle$. Here are the actions of all the 6 terms of the
Hamiltonian constraint restricted to the interval containing $v$:
\begin{eqnarray}
\hat H_K^{\theta} |k_v^{\pm}, \mu_v, \nu_v\rangle & = &
\f{\sqrt{\gamma\lP^2} }{4 \gamma^2 \mu_0\nu_0}
\Bigg[\sqrt{|\mu_v||\nu_v|}\bigg( \sqrt{| k_v^+ + k_v^- +1|} - \sqrt{|
k_v^+ +k_v^- - 1|}\bigg) \times \\ 
& & \hspace{5.4cm} \sin(\mu_0 X(\bar \theta_i)) \sin(\nu_0 Y(\bar
\theta_i)) \Bigg] |k_v^{\pm}, \mu_v, \nu_v\rangle
\nonumber \label{actionHtheta} \\
\hat H_K^{x} |k_v^{\pm}, \mu_v, \nu_v\rangle & = & 
\f{\sqrt{\gamma\lP^2} }{4 \gamma^2 \mu_0\nu_0} \Bigg[\sqrt{|k_v^+ +
k_v^-||\nu_v|}\bigg( \sqrt{| \mu_v + \mu_0|} - \sqrt{| \mu_v  -
\mu_0|}\bigg) \times \\ 
& & \hspace{0.0cm}2 \sin \left (\f{1}{2} \nu_0 Y(\bar \theta_i +
\epsilon) \right)\cos \left(\f{1}{2} \nu_0 Y(\bar \theta_{i})\right)\sin
\left( \int_{\mathcal{I}_i} {\cal A} -\Delta_i\right)\Bigg] |k_v^{\pm},
\mu_v, \nu_v\rangle 
\nonumber \label{actionHx} \\
\hat H_K^{y} |k_v^{\pm}, \mu_v, \nu_v\rangle & = & 
\f{\sqrt{\gamma\lP^2} }{4 \gamma^2 \mu_0\nu_0} \Bigg[\sqrt{|k_v^+ +
k_v^-||\mu_v|}\bigg( \sqrt{| \nu_v + \nu_0|} - \sqrt{| \nu_v  -
\nu_0|}\bigg) \times \\ 
& & \hspace{0.0cm}2 \sin \left (\f{1}{2} \nu_0 X(\bar \theta_i +
\epsilon) \right)\cos \left(\f{1}{2} \nu_0 X(\bar \theta_{i})\right)\sin
\left( \int_{\mathcal{I}_i} {\cal A} -\Delta_i\right)\Bigg] |k_v^{\pm},
\mu_v, \nu_v\rangle 
\nonumber \label{actionHy} \\
\hat H_P^{\rm (1)} |k^{\pm}_v, \mu_v, \nu_v\rangle & = & 
\left[\f{\sqrt{\gamma\lP^2}}{2}\left(
\f{1}{8\mu_0\nu_0\alpha^3}\right)^k \right]
\Bigg\{(k_v^+ - k_v^-)\Bigg\}^2 \times \nonumber \\
& & \hspace{0.0cm} \Bigg[
\left\{ \left(|\mu_v + \mu_0|^{\alpha} - |\mu_v - \mu_0|^{\alpha}\right)
|\nu_v|^{\alpha} |k_v^+ + k_v^-|^{\alpha} \right\} \times  \\
& & \hspace{0.0cm}
\left\{ |\mu_v|^{\alpha} \left( |\nu_v + \nu_0|^{\alpha} - |\nu_v +
\nu_0|^{\alpha} \right) |k_v^+ + k_v^-|^{\alpha} \right\} \times
\nonumber\\
& & \hspace{0.0cm} 
\left\{ |\mu_v|^{\alpha} |\nu_v|^{\alpha} \left( |k_v^+ + k_v^- +
1|^{\alpha} - |k_v^+ + k_v^- + 1|^{\alpha}\right) \right\} \Bigg]^k
|k^{\pm}_v, \mu_v, \nu_v\rangle \nonumber \label{actionHP1}\\
\hat H_P^{\rm (2)} |k^{\pm}_v, \mu_v, \nu_v\rangle & = & 
\left[- \f{\sqrt{\gamma\lP^2}}{2}\left(
\f{1}{8\mu_0\nu_0\alpha^3}\right)^k \right]
\Bigg\{ (k_v^+ + k_v^-)^2 \times \nonumber \\
& & \hspace{4.5cm} \left( \widehat{{\cal F}^{-1}_{x,{\cal I}_i}} \hat
{\cal F}_{x,{\cal I}_i} (\mu_v) \right)^2 \left( \widehat{{\cal
F}^{-1}_{y,{\cal I}_i}} \hat {\cal F}_{y,{\cal I}_i} (\nu_v) \right)^2
\Bigg\} \times \nonumber \\
& & \hspace{0.0cm}
\Bigg[ \left\{ \left(|\mu_v + \mu_0|^{\alpha} - |\mu_v -
\mu_0|^{\alpha}\right) |\nu_v|^{\alpha} |k_v^+ + k_v^-|^{\alpha}
\right\} \times   \\
& & \hspace{0.0cm}
\left\{ |\mu_v|^{\alpha} \left( |\nu_v + \nu_0|^{\alpha} - |\nu_v +
\nu_0|^{\alpha} \right) |k_v^+ + k_v^-|^{\alpha} \right\} \times
\nonumber \\
& & \hspace{0.0cm} 
\left\{ |\mu_v|^{\alpha} |\nu_v|^{\alpha} \left( |k_v^+ + k_v^- +
1|^{\alpha} - |k_v^+ + k_v^- + 1|^{\alpha}\right) \right\} \Bigg]^k
|k^{\pm}_v, \mu_v, \nu_v\rangle \nonumber\label{actionHP2} \\
\hat H_P^{(3)} |k^{\pm}_v, \mu_v, \nu_v\rangle & = & 
\left[- 2 \sqrt{\gamma\lP^2} \left( \f{1}{8 \mu_0\nu_0\alpha^3}\right)^k
\right]
\Bigg\{ - k_v^-( k_v^+ - k_v^- ) \Bigg\} \times \nonumber \\
& & \hspace{0.0cm}
\Bigg[ \left\{ \left(|\mu_v + \mu_0|^{\alpha} - |\mu_v -
\mu_0|^{\alpha}\right) |\nu_v|^{\alpha} |k_v^+ + k_v^-|^{\alpha}
\right\} \times   \\
& & \hspace{0.0cm}
\left\{ |\mu_v|^{\alpha} \left( |\nu_v + \nu_0|^{\alpha} - |\nu_v +
\nu_0|^{\alpha} \right) |k_v^+ + k_v^-|^{\alpha} \right\} \times
\nonumber \\
& & \hspace{0.0cm} 
\left\{ |\mu_v|^{\alpha} |\nu_v|^{\alpha} \left( |k_v^+ + k_v^- +
1|^{\alpha} - |k_v^+ + k_v^- + 1|^{\alpha}\right) \right\} \Bigg]^k
|k^{\pm}_v, \mu_v, \nu_v\rangle \nonumber\label{actionHP3} 
\end{eqnarray}
In the above, factors of $N(\bar{\theta})$ are suppressed. 

In the first three equations, we have explicitly evaluated only the
action of the diagonal operators and kept the holonomies which ``create
new vertices'' as operators acting on $|k_v^{\pm}, \mu_v, \nu_v\rangle$.
In the terms involving $\alpha$, we have to use $\alpha = \Case{2}{3} -
\Case{1}{3k}$. The last square brackets in the last three terms is the
action of the ${\cal Z}_{\alpha}({\cal I}_i)$ after the dimensional and
numerical factors are collected together in the first square bracket.
In $H_P^{(2)}$, the products of inverse flux and flux operators approach
1 only for large values of $\mu_v, \nu_v$ while for smaller values,
these products vanish.

The above actions have to be summed over all the vertices of the graph.
These being finite, the action is finite as noted before. There is no
explicit appearance of $\epsilon$. Reference to cells enclosing the
vertices (eg $\bar{\theta_i}, {\cal I}_i$), will again transfer only to
the vertices in the limit of infinite refinement. The technical issue of
limiting operator on Cyl$^*$ can be done in the same manner as in the
full theory eg as in \cite{ALReview}.

The above definitions of the quantization of the Hamiltonian constraint
constitute {\em a choice} and there are many choices possible. There is
also the issue related to ``local degrees of freedom''.  In the next
section, a preliminary discussion of these features is presented.

\section{Discussion}

Let us quickly recapitulate where we made various choices. To begin with, we
made a cell decomposition with the understanding of taking the limit of
infinitely many cells. At this stage, no reference to any state or graph
is made. In the regularization of the kinetic term we used the `inverse
volume' and `plaquette holonomies'. We could have introduced inverse
flux operators and $\hat {\cal E}$ operators to replace $1/\sqrt{E}$ and
also replaced the $X, Y, \int_{ {\cal I}_i}{\cal A}$ by
$\sin(\mu_0X)/\mu_0$ and similarly for the others. Such a replacement
would still give the classical expression back, in the limit of small
$X, Y, \epsilon$. The quantum operator however would be different. This
procedure will also deviate from the full theory. From the point of view
of the reduced theory, this is an ambiguity. Also, in the transcription
of $H_P$ in terms of holonomies and fluxes certain choices have been
made. For example, the second term in the $H_P$, could have been
manipulated in terms of inverse powers of $\sqrt{E}$ instead of
introducing inverse flux operators (eg by replacing $1/E^x = {\cal E}
E^y/(\sqrt{E})^2)$). This would lead to ${\cal E}^2\left[{\cal F}_{x,
{\cal I}_i}{\cal F}_{y, {\cal I}_{i + 1}}  - {\cal F}_{y, {\cal
I}_i}{\cal F}_{x, {\cal I}_{i + 1}}\right]^2$ and lead to $\alpha(k) =
2/3 - 5/(3k)$. In the limit of infinite refinement, each cell will
contain {\em at most} one vertex and the cells adjacent to such a cell
will always be empty. Consequently, the second term of $H_P$, regulated
in the above manner will always give a zero action. Over and above these
different transcriptions, we also have the ambiguity introduced by the
arbitrary positive power $k$ (and $\alpha(k)$) as well as that
introduced by the arbitrary power $l \in (0, 1)$ in the definition of
inverse flux operators. All these ambiguities refer to the transcription
stage.

There are also issues related to the choice of partitions, subsequent
$\epsilon \to 0$ limit and the presence/absence of local degrees of
freedom. This is most dramatically brought out by the second term of
$H_P$. Classically, this is the term which reveals spatial correlations
in a solution space-time through $\partial_{\theta} \ln (E^y/E^x)$
\cite{GowdyClassical} and reflect the infinitely many, physical
solutions. In the (vacuum) spherically symmetric case, such a term is
absent and so are local physical degrees of freedom. We would like to
see if there is a quantization of this term which reflects these
correlations. The quantization chosen above does not correlate $\mu,
\nu$ labels at different vertices. 

In general, given a graph, a partition may be chosen to have (i) every
cell containing {\em at least} one vertex or (ii) every cell containing
{\em exactly} one vertex or (iii) every cell containing {\em at most}
one vertex. In this classification, we assume that a vertex is never a
boundary-point of a cell, which is always possible to choose.  Infinite
refinement is possible only for (iii) which we have been assuming so
far. This is the reason that in the contribution from the ${\cal I}_i$
cell, the terms referring to ${\cal I}_{i + 1}$ drop out. 

We could introduce a fourth case by requiring; (iv) every vertex to be a
boundary point of a cell. Then we would receive contributions from two
adjacent cells.  However, in $H_P^{(2)}$ (eq.
\ref{SecondTermDetailed}), the two terms with labels ${\cal I}_{i + 1},
{\cal I}_i$, both give equal contribution such that the total is zero!
The same would happen in the alternative expression given above. It
seems that in either of (iii) or (iv) type partitions, we will either
get a zero or a contribution depending only on a single vertex. Note
that these are the only partitions which allow infinite refinement
($\epsilon \to 0$) in a diffeo-covariant manner. 

We can give up on the infinite partitions (and $\epsilon \to 0$ limit)
and consider instead case (ii) partitions -- each cell contains exactly
one vertex (say in the interior). Now the contributions will explicitly
depend upon $\mu, \nu$ labels of adjacent vertices and in this sense,
spatial correlations will survive in the constraint operator. An even
more restrictive choice would be to choose the partition defined by the
graph itself - cells defined by the edges and the boundary points of
cells as vertices. In this case, the new vertices created by $H_K$ would
be the already present vertices and the constraint equation would lead
to a (partial) difference equation among the labels. This case has been
considered in the spherical symmetry \cite{Spherical2} and corresponds
to `effective operator viewpoint' discussed by Thiemann in
\cite{LQGRev}. The $\epsilon \to 0$ limit may then be thought to be
relevant when states have support on graphs with very large (but finite)
number vertices, heuristically for semiclassical states.  Whether
requiring the constraint algebra to be satisfied on diffeomorphism
invariant states chooses/restricts the alternatives and ambiguities
remains to be seen and will be explored in the third paper in this
series. 

It is important to be able to identify the quantum theory of the
symmetry reduced model as a `sector' of the full theory. There are more
than one ways for such an identification \cite{Engle} and this is still
an open problem. The midi-superspace (inhomogeneous) model is in between
the full theory and the mini-superspace (homogeneous) models. This
provides an opportunity to explore the identification of the appropriate
homogeneous (anisotropic as well as isotropic) models as sectors of the
Gowdy model\footnote{ We thank an anonymous referee for pointing this
out.}. However, at present, we do not have any specific results to
report.

\begin{acknowledgments} 
Discussions with Alok Laddha are gratefully acknowledged.
\end{acknowledgments}


\begin{thebibliography}{10}

\bibitem{LQGRev} Rovelli C, Loop Quantum Gravity, 1998 Living Reviews in
Relativity {\bf 1}, 1, [gr-qc/9710008]; \\
%
Thiemann T, Introduction to Modern Canonical Quantum General Relativity, 

[gr-qc/0110034].
%
\bibitem{ALReview}
Ashtekar A, and Lewandowski J, 2004
Background Independent Quantum Gravity: A Status Report,
{\em Class. Quant. Grav.}, {\bf 21}, R53,
[gr-qc/0404018].
%
\bibitem{LQCRev}
Bojowald M, 2005, Loop Quantum Cosmology, {\em Living Rev. Rel.}, 
{\bf 8}, 11, [gr-qc/0601085]. 
%
\bibitem{Gowdy}
Gowdy R H, 1974 Vacuum space-times with two parameter spacelike
isometry groups and compact invariant hypersurfaces: Topologies and
boundary conditions, {\em Ann. Phys.} {\bf 83} 203-241.
%

\bibitem{ClassGowdy}
Moncrief V, 1981
Infinite-dimensional family of vacuum
cosmological models with Taub-NUT (Newman-Unti-Tamburino)-type
extensions {\em Phys. Rev} {\bf D 23} 312-315; \\
%
Isenberg J and Moncrief V, 1990
Asymptotic behavior of the
gravitational field and the nature of singularities in Gowdy space-time
{\em Ann. Phys.} {\bf 199} 84-122.
%
\bibitem{ADMQuantization}
Misner C W,  1973 A Minisuperspace Example: The Gowdy $T^{3}$ Cosmology
{\em Phys. Rev.} {\bf D 8} 3271-3285;
\\%
Berger B K, 1975 Quantum cosmology: Exact solution for the Gowdy $T^{3}$ model
{\em Phys. Rev.} {\bf D 11} 2770-2780.
%

\bibitem{Pierri}
Pierri M,  2002  Probing quantum general relativity through exactly 
soluble midi-superspaces. II: Polarized Gowdy models
{\em Int. J. Mod. Phys.} {\bf D11} 135,
[gr-qc/0101013].
%

\bibitem{MenamaruganTorre}
Corichi A, Cortez J and Quevedo H, 2002
On unitary time evolution in Gowdy $T^3$ cosmologies
{\em Int. J. Mod. Phys.} {\bf D 11} 1451-1468
[gr-qc/0204053];
\\%
Torre C G, 2002  Quantum dynamics of the polarized Gowdy $T^3$ model
{\em Phys. Rev.} {\bf D 66} 084017
[gr-qc/0206083];
\\%
Torre C G, 2006  Observables for the polarized Gowdy model
{\em Class. Quant. Grav.} {\bf 23} 1543-1556
[gr-qc/0508008];
\\%
Torre C G, 2007  Schroedinger representation for the polarized Gowdy model
{\em Class. Quant. Grav.} {\bf 24} 1-13 [gr-qc/0607084];
\\%
Cortez J and Mena Marugan G A,  2005
Feasibility of a unitary quantum dynamics in the Gowdy $T^3$ cosmological model
{\em Phys. Rev.} {\bf D 72} 064020
[gr-qc/0507139].
%

\bibitem{Corichi}
Corichi A, Cortez J and Mena Marugan G A,  2006
Unitary evolution in Gowdy cosmology
{\em Phys. Rev.} {\bf D 73} 041502
[gr-qc/0510109];
\\%
Corichi A, Cortez J and Mena Marugan G A,  2006
Quantum Gowdy $T^3$ model: A unitary description
{\em Phys. Rev.} {\bf D 73}  084020
[gr-qc/0603006];
\\%
Corichi A, Cortez J, Mena Marugan G A and Velhinho J M,  2006
Quantum Gowdy $T^3$ model: A uniqueness result
{\em Class. Quant. Grav.}  {\bf 23} 6301
[gr-qc/0607136];
\\%
Cortez J, Mena Marugan G A and Velhinho J M,  2007
Uniqueness of the Fock quantization of the Gowdy $T^3$ model
{\em Phys. Rev.} {\bf D 75} 084027
[gr-qc/0702117];
\\%
Corichi A, Cortez J, Mena Marugan G A and Velhinho J M,  2007
Quantum Gowdy $T^3$ Model: Schrodinger Representation with Unitary Dynamics
{\em Phys. Rev.}  {\bf D 76} 124031
[arXiv:0710.0277].

\bibitem{HusainSmolin}
%
Husain V and Smolin L, 1989 Exactly Soluble Quantum Cosmologies
from Two Killing Field Reductions of General Relativity {\em Nucl.
Phys.} {\bf B 327} 205-238.

\bibitem{Menamarugan}
%
Mena Marugan G A, 1997 Canonical quantization of the Gowdy model
{\em Phys. Rev.} {\bf D 56} 908-919, [gr-qc/9704041].

\bibitem{MGM}
%
Martin-Benito M and Garay L J and Mena Marugan G A, 2008
Hybrid Quantum Gowdy Cosmology: Combining Loop and Fock Quantizations,
arXiv:0804.1098 [gr-qc]

\bibitem{Spherical1}
Bojowald M, 2004
Spherically symmetric quantum geometry: States and basic operators,
{\em Class. Quant. Grav.} {\bf 21}, 3733-3753,
[gr-qc/0407017].
%
\bibitem{Spherical2}
Bojowald M and Swiderski R, 2006
Spherically Symmetric Quantum Geometry: Hamiltonian Constraint,
{\em Class. Quant. Grav.} {\bf 23}, 2129-2154,
[gr-qc/0511108].
%
\bibitem{GowdyClassical}
Banerjee K and Date G, 2008
Loop Quantization of Polarized Gowdy Model on $T^3$: Classical Theory,
{\em Class. Quantum Grav.} {\bf 25} 105014,
arXiv:0712.0683 [gr-qc].
%
\bibitem{QSD5}
Thiemann T, 1998
QSD V: Quantum gravity as the natural regulator of matter quantum field
theories,
{\em Class. Quant. Grav.} {\bf 15}, 1281-1314,
[gr-qc/9705019].
%
\bibitem{InverseTriad}
Bojowald M, 2001
The inverse scale factor in isotropic quantum geometry,
{\em Phys. Rev.} {\bf D 64}, 084018,
[gr-qc/0105067].
%
\bibitem{Engle}
Bojowald M and Kastrup H A, 2000
Quantum symmetry reduction for diffeomorphism invariant theories of connections,
{\em Class. Quant. Grav.} {\bf 17}, 3009,
[hep-th/9907042];
\\%
Engle J, 2006
Quantum field theory and its symmetry reduction,
{\em Class. Quant. Grav.}  {\bf 23}, 2861,
[gr-qc/0511107];
\\%
Engle J, 2007
Relating loop quantum cosmology to loop quantum gravity: Symmetric sectors 
and embeddings, {\em  Class. Quantum Grav.} {\bf 24} 5777-5802
[gr-qc/0701132].
%
\end{thebibliography}
\end{document}